\documentclass[sigconf,authorversion]{acmart}

\usepackage[utf8]{inputenc} 	
\usepackage{booktabs}			
\usepackage{fontawesome}    	
\usepackage{tikz}				
\usepackage{colortbl,xcolor} 	

\newcommand{\male}{\faMars}
\newcommand{\female}{\faVenus}
\definecolor{mygrey}{rgb}{0.9, 0.9, 0.9}

\copyrightyear{2018} 
\acmYear{2018} 
\setcopyright{acmlicensed}
\acmConference[SIGCSE '18]{SIGCSE '18: The 49th ACM Technical Symposium on Computing Science Education}{February 21--24, 2018}{Baltimore, MD, USA}
\acmPrice{15.00}
\acmDOI{10.1145/3159450.3159454}
\acmISBN{978-1-4503-5103-4/18/02}

\begin{document}
\title[Challenges Arising from Prerequisite Testing in Cybersecurity Games]{Challenges Arising from Prerequisite Testing in~Cybersecurity~Games}

\author{Valdemar Švábenský}
\orcid{0000-0001-8546-280X}
\affiliation{
  \institution{Masaryk University}
  \department{Faculty of Informatics}
}
\affiliation{
  \department{Institute of Computer Science}
  \streetaddress{Botanická 68a}
  \city{Brno} 
  \state{Czech Republic}
  \postcode{60200}
}
\email{svabensky@ics.muni.cz}

\author{Jan Vykopal}
\orcid{0000-0002-3425-0951}
\affiliation{%
  \institution{Masaryk University}
  \department{Institute of Computer Science}
  \streetaddress{Botanická 68a}
  \city{Brno} 
  \state{Czech Republic}
  \postcode{60200}
}
\email{vykopal@ics.muni.cz}

\begin{abstract}
Cybersecurity games are an attractive and popular method of active learning. However, the majority of current games are created for advanced players, which often leads to frustration in less experienced learners. Therefore, we decided to focus on a diagnostic assessment of participants entering the games. We assume that information about the players' knowledge, skills, and experience enables tutors or learning environments to suitably assist participants with game challenges and maximize learning in their virtual adventure. In this paper, we present a pioneering experiment examining the predictive value of a short quiz and self-assessment for identifying learners' readiness before playing a cybersecurity game. We hypothesized that these predictors would model players' performance. A linear regression analysis showed that the game performance can be accurately predicted by well-designed prerequisite testing, but not by self-assessment. At the same time, we identified major challenges related to the design of pretests for cybersecurity games: calibrating test questions with respect to the skills relevant for the game, minimizing the quiz's length while maximizing its informative value, and embedding the pretest in the game. Our results are relevant for educational researchers and cybersecurity instructors of students at all learning levels.
\end{abstract}



\maketitle

\section{Introduction}

Cybersecurity games allow participants to test their knowledge and exercise their skills in different areas of computer security. Although carried out in a closed and controlled environment, the games simulate practical, real-world situations. The players can attack and defend computer systems, analyze network traffic, or disassemble binaries without any negative consequences in reality.

Studies confirm multiple benefits of cybersecurity games~\cite{pusey, tobey, wouters}. They can inspire interest in computer security and motivate participants to explore the field further. Games designed specifically for education enrich the curriculum and test the learners' competence in an authentic setting, enabling them to discover their strengths and weaknesses. Ranking well in competitive games often leads to peer recognition, (monetary) prizes, or job opportunities.

Competitions and games of various difficulty levels and focus are spreading widely, from informal online hacking communities to universities and professional security conferences. The number of participants in cybersecurity games is growing exponentially~\cite{tobey}. At the same time, several authors argue that although high-quality games are available, they offer little educational value to learners~\cite{pusey, werther}. The games often require substantial knowledge of the problem domain, as well as practical expertise, in advance. As a result, the majority of computer science students are unable to participate. Even worse, some students' interest and motivation may diminish after an unsuccessful attempt \cite{pusey}. Research suggests that games and contests are effective only for already skilled players, whose skills ``closely match those required by the competition''~\cite{tobey}.

Achieving \textit{game balance} (assigning tasks that are suitable for the player's skill, neither trivial nor impossible to solve~\cite{pusey, nagarajan}) is vital in educational games. One approach to achieving game balance is introducing methods of adaptive learning~\cite{gondree}, which change the difficulty of the tasks during the game based on the player's success rate. Another solution is a diagnostic assessment by prerequisite testing, which is the topic of this paper. This approach, suggested in pedagogical theory~\cite{morrison, kommers}, refers to testing the player before or during a game to determine whether the player's skills are sufficient to finish the tasks, thus providing game balance~\cite{pusey}.

This work's main motivation is the demand for timely identification of students who may require help while playing, so that their individual needs can be appropriately addressed. This can be done by providing learners with more precise instructions, hints, or relevant study materials.

To explore the predictive value of both self-assessment and prerequisite tests, we conducted an experiment involving 67 learners and two games. They completed a short questionnaire, played a game, and then reflected on their experience in another survey. Tracking the players' actions during the game allows us to compare the assessment results with the in-game performance. Based on this experiment, we seek to answer two research questions:

\vspace*{-0.03cm}
\begin{enumerate}
	\item Can a prerequisite test accurately model learners' performance in cybersecurity games?
	\item Is self-assessment a reliable indicator of ability in the context of cybersecurity?
\end{enumerate}


\vspace*{-0.2cm}
\section{Related work} \label{sec:related}

Research into diagnostic assessment of cybersecurity skills is sparse and not yet mature. Therefore, this section also mentions works dealing with assessment and testing in other 
educational
domains. 

While Nagarajan et al.~\cite{nagarajan} stress that measuring skills before and after playing is vital to determine the game's effectiveness, they report that security training programs do not implement this measurement. To the best of our knowledge, we can confirm this observation.

Mirkovic et al.~\cite{mirkovic, 3gse} emphasize considering individual skills in team cybersecurity games to balance the teams and give everyone an equal chance to succeed. Before using cybersecurity games in classrooms, Mirkovic and Peterson~\cite{mirkovic} surveyed the students about their skills to create balanced teams. Unfortunately, the paper does not provide details about the process. In another study~\cite{3gse}, the participants reported their knowledge of programming, security, and tools that were to be used later. Again, the survey results were used to balance the teams. However, the authors concluded that this led to inequality among the teams, as the self-assessment was often inaccurate. They recommend ``conducting a short quiz-type assessment prior to the event'', but do not specify how to do this.

Next, Bolívar-Cruz et al.~\cite{bolivar} examined self-assessment of university students in oral communication. Their literature review shows that self-assessment's accuracy is generally low or questionable, but also warns readers about methodological errors in some of the previous studies. Allen and Van Der Velden~\cite{allen} advocate using self-assessment complemented by independent, objective tests to increase the reliability of results. They argue that people know the level of their skills best but also warn readers of its issues, including misunderstood skill items, an ambiguous rating scale, and the risk of an unreliable answer (either intentional or not).

Finally, Govindasamy~\cite{govindasamy} suggests applying pretests in e-learning courses to test both minimum requirements and proficiency. Based on the results, the learner can be directed to a simpler or more difficult course, or skip the already familiar areas in the current course. Educational literature~\cite{petty, morrison, kommers} advocates the use of prerequisite testing in teaching practice.

\section{Structure of a cybersecurity game} \label{sec:game}

We use instances of a cybersecurity game following a generic format of a hands-on activity, which is performed in a realistic network environment emulated by the KYPO cyber range \cite{kypo2017}.

\begin{figure}
\centering
\includegraphics[width=0.9\linewidth]{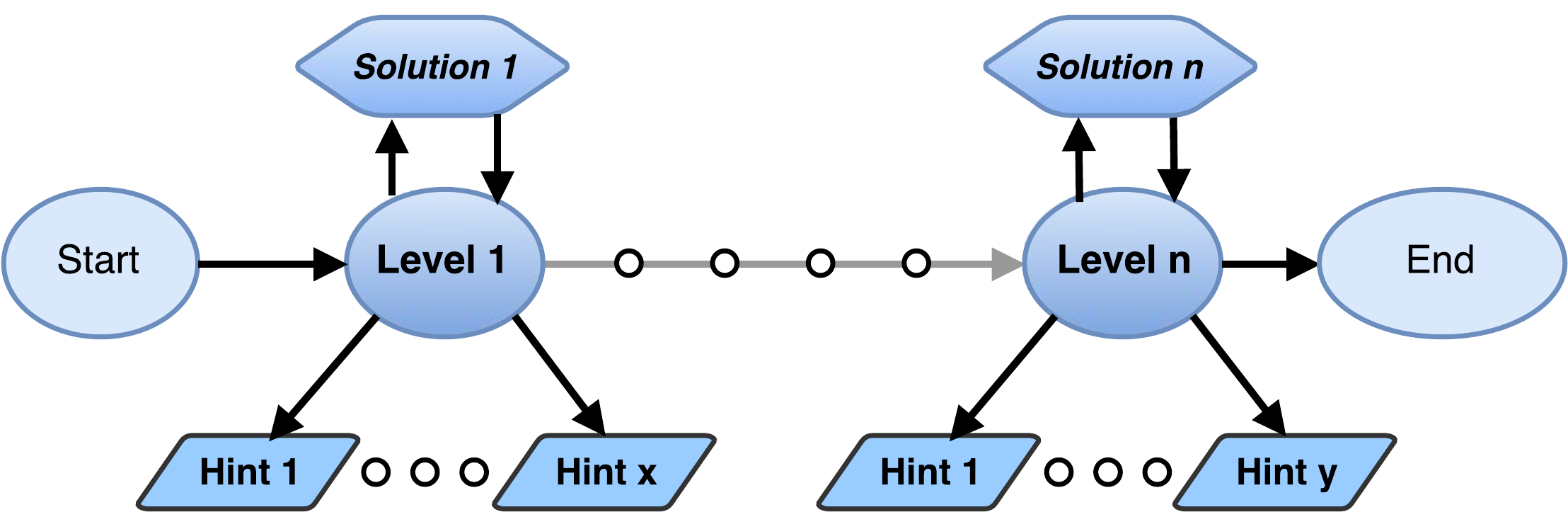}
\caption{The general structure of the cybersecurity game used in the experiment}
\label{fig:ctf-scheme}
\end{figure}

Figure~\ref{fig:ctf-scheme} shows the scheme of the game, which is structured into successive levels leading to the final objective, such as data theft. Before the start, each player has access to limited network resources and a brief information about the goal. Every level is finished by finding a correct flag (a short string); this accomplishment is awarded a specified number of points contributing to the player’s total score. The game ends upon entering the last flag or when a predefined final check of the system’s state succeeds.

The game provides optional scaffolding by offering hints. If the player struggles with a level, these hints can be used in exchange for penalization by negative points. There is evidence that game elements such as points and levels can improve the overall effectiveness of learning~\cite{jang}. It is also possible to skip the level, display the recommended solution, and quit the game at any time.

The generic nature of the game format allows us to collect generic game events, regardless of the topic of the particular game and technical infrastructure used. The game events describe the player's interaction with the game interface, namely: starting and ending the game or each level, submitting incorrect flags and their content, using hints, skipping a level, and displaying a solution. Each event contains a timestamp and a unique ID of the player.

\section{Experiment Design} \label{sec:experiment}

Table \ref{table:experiment-design} summarizes the phases and structure of the experiment. At first, each player completed a prerequisite test and a self-assessment questionnaire. The players then proceeded to a game, where their performance was tracked using the generic game events. Finally, the players filled in a post-game feedback questionnaire. The self-assessment and pretest data were used to create a linear regression model of learners' skill, which is expressed by two metrics: the total game score and number of levels finished.

\subsection{Participants}

\begin{table}
\begin{center}
\caption{The design of the experiment with the number of participants and their gender}
\vspace{-0.2cm}
\begin{tabular}{|c|c|c|c|c|}
\hline
\multicolumn{5}{|c|}{67 participants}                          \\ \hline
Session 1  &  Session 2 & Session 3  & Session 4  & \cellcolor{mygrey} Session 5  \\ 
9 \male, 1 \female & 4 \male, 3 \female & 15 \male, 3 \female & 12 \male, 0 \female & \cellcolor{mygrey} 19 \male, 1 \female \\ \hline \hline
\multicolumn{4}{|c|}{Pretest 1} & \cellcolor{mygrey} Pretest 2                    \\ 
\multicolumn{3}{|c|}{\qquad\qquad\quad 5 items} & + 2 items & \cellcolor{mygrey} 4 items         \\ \hline
 \multicolumn{5}{|c|}{Self-assessment}                         \\
 \multicolumn{5}{|c|}{3 items}
\\ \hline
\multicolumn{4}{|c|}{Game 1} & \cellcolor{mygrey} Game 2                          \\
\multicolumn{4}{|c|}{6 levels} & \cellcolor{mygrey} 4 levels                      \\ \hline
\multicolumn{5}{|c|}{Post-game feedback}                       \\ \hline
\end{tabular}
\label{table:experiment-design}
\end{center}
\vspace*{-0.5cm}
\end{table}

A total of 67 cybersecurity students and professionals of various levels of expertise, background, and nationality participated, covering a broad spectrum of the games' target audience. The players' only motivation was their interest, as they did not receive any incentives for taking part in the study. The participants were informed about the intended use of the acquired data solely for the purpose of this experiment. The data was anonymized during the processing.

The learners were divided into five game sessions, each lasting two hours. The first session included 10 computer science students from St. Pölten University of Applied Sciences, Austria. The second session consisted of 7 employees of the Computer Security Incident Response Team of Pavol Jozef Šafárik University in Košice, Slovakia. The third session included 18 computer science students from Masaryk University, Brno, Czech Republic. The fourth session included 12 finalists of a Czech high school cybersecurity competition. The fifth session included 20 attendees of the AIMS 2017 conference on network security and management, held in Zurich, Switzerland.

\subsection{Selected cybersecurity games}

We have two games G1 and G2 in this experiment to test various game tasks and prerequisite tests. Table~\ref{table:game-time-score} details the maximum score, the number of hints, and scoring penalties for taking the hints in each level of both games.

The topic of the G1 game played in the first four game sessions is information theft from a database server of a fictitious bank. Each player initially controls a single Linux host in an unknown network. The player must gradually gain and maintain access to other hosts that are a part of the bank's network infrastructure, and, finally, steal confidential information. This mission is split into six levels, in which the players exercise penetration testing skills.

To provide comparative data, 20 participants of the fifth game session played another game, G2, with the topic of gaining access to a remote server and destroying stored data. This game is split into four levels with learning objectives similar to the levels of G1.

\begin{table}
\begin{center}
\caption{Characteristics of the selected games G1 and G2}
\begin{tabular}{|l|r|r|r|r|r|r|r|r|r|r|}
\hline 
Level & \multicolumn{2}{c}{Score [pts]} & \multicolumn{2}{|c}{Hints} & \multicolumn{2}{|c|}{Penalty [pts]} \\ \hline
      &      G1 &  G2 &     G1 &  G2 &                   G1 & G2                      \\ \hline \hline
    1 &       8 &  16 &      2 &   2 &           $-2$, $-2$ &              $-3$, $-5$ \\ \hline 
    2 &      12 &  22 &      2 &   2 &           $-2$, $-3$ &              $-7$, $-5$ \\ \hline 
    3 &      23 &  27 &      3 &   4 &     $-1$, $-3$, $-2$ & $-5$, $-10$, $-0$, $-0$ \\ \hline 
    4 &      20 &  35 &      2 &   4 &           $-2$, $-3$ & $-5$, $-10$, $-5$, $-5$ \\ \hline 
    5 &      22 & --- &      2 & --- &           $-3$, $-4$ &                     --- \\ \hline 
    6 &      15 & --- &      2 & --- &           $-5$, $-2$ &                     --- \\ \hline \hline
Total &     100 & 100 &     13 &  12 &                $-34$ &                   $-60$ \\ \hline 
\end{tabular}
\label{table:game-time-score}
\end{center}
\end{table}

\subsection{Skill measurement before the game}

We created quizzes testing prerequisites for each of the two selected games using a model for question design by Beatty et al.~\cite{beatty}. Details about the pretest's design can be found in~\cite{diplomka}. In the first three game sessions, the players completed a pretest consisting of five questions. In the fourth session, this pretest was enhanced by two extra questions. In the fifth session, a different but similar pretest consisting of four questions was used. The prerequisite quizzes capture a representative sample of key knowledge and skills exercised in the games. Below is an example of a question for the first level of both games (an asterisk marks the correct answer): 

\begin{footnotesize}
\begin{verbatim}
What is the effect of the command ping 10.0.0.3?
a)* Tests the reachability of a host with an IP address 10.0.0.3.
b)  Scans open ports of the server with an IP address 10.0.0.3.
c)  Error, the syntax of the command is incorrect.
d)  Measures the number of network hops to a host with an IP address 
    10.0.0.3.
\end{verbatim}
\end{footnotesize}

To evaluate the test, we used a simple \textit{dichotomous} scoring method awarding one point for a fully correct answer, and zero points for a partially or entirely incorrect answer per question. Moreover, after responding to each question, a learner rated the level of certainty in the answer on a five-step scale developed by Hassmén and Hunt~\cite{hassmen, hunt}. This scoring method, referred to as \textit{confidence assessment}, yielded another test score. In both cases, a sum of the respective scores was considered as an estimate of each learner's total readiness.

Apart from the prerequisite test, each player completed a 3-item self-assessment questionnaire before starting the game. Since there is no standardized methodology for designing the questions,
we created them with respect to the content of the particular games. The survey asked the players to self-evaluate their expertise with using three tools needed in the games: for port scanning (Nmap), vulnerability exploiting (Metasploit), and password attacks (John the Ripper). For each tool, the player selected one of four levels of competence on the following ordered scale: zero experience, beginner (basic knowledge), intermediate (some practical experience), and expert (professional working experience). To aggregate the learners' input, we used the median to express the central tendency of each player's self-assessment. Since the self-assessment data are ordinal, we avoided using the arithmetic mean.

\subsection{Post-game feedback}

After finishing the game, the participants completed a post-game feedback questionnaire. The goal of the survey was to have each player subjectively assess the game's difficulty on a scale from 1 (trivial) to 5 (impossible), and reflect if any learning occurred. This reflection helps to determine if game balance was achieved, and if the player perceived the game as educational. Unfortunately, due to a technical error, we collected the results from only 46 participants.

\section{Results} \label{sec:results}

\begin{table*}
\centering
\begin{minipage}[b]{.5\textwidth}
  \caption{Examined variables and descriptive statistics of the participant data for the four G1 sessions and 5-item pretest}    
  \vspace{-0.2cm}
  \begin{tabular}{|*{2}{l|}l|r|r|r|r|r|}
  \hline
  \multicolumn{2}{|l|}{Variable} & Possible range & Min & Max & Avg & Med \\ \hline
  \hline
  Self-assessment  & $S$   & 0 to 3        & 0     & 2   & $0.9$   & 1   \\ \hline
  Pretest (dich.)  & $P_d$ & 0 to 5        & 0     & 5   & $3.9$   & 4   \\ \hline
  Pretest (conf.)  & $P_c$ & $-300$ to 250 & $-91$ & 240 & $150.6$ & 155 \\ \hline
  Game score       & $T$   & 0 to 100      & 0     & 100 & $49.7$  & 55  \\ \hline
  Levels finished  & $L$   & 0 to 6        & 0     & 6   & $3.5$   & 4   \\ \hline
  Difficulty       & $D$   & 1 to 5        & 2     & 5   & $3.5$   & 3   \\ \hline
  \end{tabular}
  \label{table:stats}
\end{minipage}%
\hfill
\begin{minipage}[b]{.45\textwidth}
  \caption{The overall best linear regression models (see Table \ref{table:stats} for the description of the variables).}
  \vspace{-0.2cm}
  \begin{tabular}{|l|l|r|r|r|}
  \hline
  Game & Model & $R^2$ & F-statistic & p-value \\ \hline
  \hline
  G1 & $T = 8.98 + 10.34 \cdot P_d$ & $0.17$ & $9.36$ &  $0.004$ \\ \hline
  G1 & $T = 10.87 + 0.23 \cdot P_c$ & $0.31$ & $15.66$ & $<0.001$ \\ \hline
  G1 & $L = 1.62 +  0.48 \cdot P_d$ & $0.14$ & $7.22$ &  $0.010$ \\ \hline
  G1 & $L = 1.73 +  0.01 \cdot P_c$ & $0.24$ & $10.99$ &  $0.002$ \\ \hline
  \end{tabular}
  \\ 
  \\
  \\
  \label{table:models}
\end{minipage}%
\end{table*}

\begin{figure*}
\centering
\vspace*{-1.8cm}
\begin{minipage}{.5\textwidth}
\begin{tikzpicture}[x=1pt,y=1pt]
\definecolor{fillColor}{RGB}{255,255,255}
\path[use as bounding box,fill=fillColor,fill opacity=0.00] (0,0) rectangle (260.17,180.67);
\begin{scope}
\path[clip] ( 49.20, 61.20) rectangle (234.97,131.47);
\definecolor{drawColor}{RGB}{0,0,0}

\path[draw=drawColor,line width= 1.2pt,line join=round] ( 63.25, 70.96) -- ( 77.58, 70.96);

\path[draw=drawColor,line width= 0.4pt,dash pattern=on 4pt off 4pt ,line join=round,line cap=round] ( 70.41, 70.96) -- ( 70.41, 70.96);

\path[draw=drawColor,line width= 0.4pt,dash pattern=on 4pt off 4pt ,line join=round,line cap=round] ( 70.41, 70.96) -- ( 70.41, 70.96);

\path[draw=drawColor,line width= 0.4pt,line join=round,line cap=round] ( 66.83, 70.96) -- ( 74.00, 70.96);

\path[draw=drawColor,line width= 0.4pt,line join=round,line cap=round] ( 66.83, 70.96) -- ( 74.00, 70.96);

\path[draw=drawColor,line width= 0.4pt,line join=round,line cap=round] ( 63.25, 70.96) --
	( 77.58, 70.96) --
	( 77.58, 70.96) --
	( 63.25, 70.96) --
	( 63.25, 70.96);

\path[draw=drawColor,line width= 1.2pt,line join=round] (120.58, 72.26) -- (134.92, 72.26);

\path[draw=drawColor,line width= 0.4pt,dash pattern=on 4pt off 4pt ,line join=round,line cap=round] (127.75, 63.80) -- (127.75, 68.03);

\path[draw=drawColor,line width= 0.4pt,dash pattern=on 4pt off 4pt ,line join=round,line cap=round] (127.75,115.21) -- (127.75, 93.73);

\path[draw=drawColor,line width= 0.4pt,line join=round,line cap=round] (124.17, 63.80) -- (131.34, 63.80);

\path[draw=drawColor,line width= 0.4pt,line join=round,line cap=round] (124.17,115.21) -- (131.34,115.21);

\path[draw=drawColor,line width= 0.4pt,line join=round,line cap=round] (120.58, 68.03) --
	(134.92, 68.03) --
	(134.92, 93.73) --
	(120.58, 93.73) --
	(120.58, 68.03);

\path[draw=drawColor,line width= 1.2pt,line join=round] (149.25, 98.29) -- (163.59, 98.29);

\path[draw=drawColor,line width= 0.4pt,dash pattern=on 4pt off 4pt ,line join=round,line cap=round] (156.42, 67.71) -- (156.42, 87.23);

\path[draw=drawColor,line width= 0.4pt,dash pattern=on 4pt off 4pt ,line join=round,line cap=round] (156.42,110.65) -- (156.42,104.80);

\path[draw=drawColor,line width= 0.4pt,line join=round,line cap=round] (152.84, 67.71) -- (160.00, 67.71);

\path[draw=drawColor,line width= 0.4pt,line join=round,line cap=round] (152.84,110.65) -- (160.00,110.65);

\path[draw=drawColor,line width= 0.4pt,line join=round,line cap=round] (149.25, 87.23) --
	(163.59, 87.23) --
	(163.59,104.80) --
	(149.25,104.80) --
	(149.25, 87.23);

\path[draw=drawColor,line width= 1.2pt,line join=round] (177.92, 90.48) -- (192.26, 90.48);

\path[draw=drawColor,line width= 0.4pt,dash pattern=on 4pt off 4pt ,line join=round,line cap=round] (185.09, 66.41) -- (185.09, 80.07);

\path[draw=drawColor,line width= 0.4pt,dash pattern=on 4pt off 4pt ,line join=round,line cap=round] (185.09,116.51) -- (185.09,103.50);

\path[draw=drawColor,line width= 0.4pt,line join=round,line cap=round] (181.51, 66.41) -- (188.67, 66.41);

\path[draw=drawColor,line width= 0.4pt,line join=round,line cap=round] (181.51,116.51) -- (188.67,116.51);

\path[draw=drawColor,line width= 0.4pt,line join=round,line cap=round] (177.92, 80.07) --
	(192.26, 80.07) --
	(192.26,103.50) --
	(177.92,103.50) --
	(177.92, 80.07);

\path[draw=drawColor,line width= 1.2pt,line join=round] (206.59,112.28) -- (220.92,112.28);

\path[draw=drawColor,line width= 0.4pt,dash pattern=on 4pt off 4pt ,line join=round,line cap=round] (213.76, 82.02) -- (213.76, 97.64);

\path[draw=drawColor,line width= 0.4pt,dash pattern=on 4pt off 4pt ,line join=round,line cap=round] (213.76,128.87) -- (213.76,116.18);

\path[draw=drawColor,line width= 0.4pt,line join=round,line cap=round] (210.17, 82.02) -- (217.34, 82.02);

\path[draw=drawColor,line width= 0.4pt,line join=round,line cap=round] (210.17,128.87) -- (217.34,128.87);

\path[draw=drawColor,line width= 0.4pt,line join=round,line cap=round] (206.59, 97.64) --
	(220.92, 97.64) --
	(220.92,116.18) --
	(206.59,116.18) --
	(206.59, 97.64);

\path[draw=drawColor,line width= 0.4pt,line join=round,line cap=round] (213.76, 69.01) circle (  2.25);
\end{scope}
\begin{scope}
\path[clip] (  0.00,  0.00) rectangle (260.17,180.67);
\definecolor{drawColor}{RGB}{0,0,0}

\path[draw=drawColor,line width= 0.4pt,line join=round,line cap=round] ( 70.41, 61.20) -- (213.76, 61.20);

\path[draw=drawColor,line width= 0.4pt,line join=round,line cap=round] ( 70.41, 61.20) -- ( 70.41, 55.20);

\path[draw=drawColor,line width= 0.4pt,line join=round,line cap=round] ( 99.08, 61.20) -- ( 99.08, 55.20);

\path[draw=drawColor,line width= 0.4pt,line join=round,line cap=round] (127.75, 61.20) -- (127.75, 55.20);

\path[draw=drawColor,line width= 0.4pt,line join=round,line cap=round] (156.42, 61.20) -- (156.42, 55.20);

\path[draw=drawColor,line width= 0.4pt,line join=round,line cap=round] (185.09, 61.20) -- (185.09, 55.20);

\path[draw=drawColor,line width= 0.4pt,line join=round,line cap=round] (213.76, 61.20) -- (213.76, 55.20);

\node[text=drawColor,anchor=base,inner sep=0pt, outer sep=0pt, scale=  1.00] at ( 70.41, 44.60) {0};

\node[text=drawColor,anchor=base,inner sep=0pt, outer sep=0pt, scale=  1.00] at ( 99.08, 44.60) {1};

\node[text=drawColor,anchor=base,inner sep=0pt, outer sep=0pt, scale=  1.00] at (127.75, 44.60) {2};

\node[text=drawColor,anchor=base,inner sep=0pt, outer sep=0pt, scale=  1.00] at (156.42, 44.60) {3};

\node[text=drawColor,anchor=base,inner sep=0pt, outer sep=0pt, scale=  1.00] at (185.09, 44.60) {4};

\node[text=drawColor,anchor=base,inner sep=0pt, outer sep=0pt, scale=  1.00] at (213.76, 44.60) {5};

\path[draw=drawColor,line width= 0.4pt,line join=round,line cap=round] ( 49.20, 63.80) -- ( 49.20,128.87);

\path[draw=drawColor,line width= 0.4pt,line join=round,line cap=round] ( 49.20, 63.80) -- ( 43.20, 63.80);

\path[draw=drawColor,line width= 0.4pt,line join=round,line cap=round] ( 49.20, 76.82) -- ( 43.20, 76.82);

\path[draw=drawColor,line width= 0.4pt,line join=round,line cap=round] ( 49.20, 89.83) -- ( 43.20, 89.83);

\path[draw=drawColor,line width= 0.4pt,line join=round,line cap=round] ( 49.20,102.84) -- ( 43.20,102.84);

\path[draw=drawColor,line width= 0.4pt,line join=round,line cap=round] ( 49.20,115.86) -- ( 43.20,115.86);

\path[draw=drawColor,line width= 0.4pt,line join=round,line cap=round] ( 49.20,128.87) -- ( 43.20,128.87);

\node[text=drawColor,anchor=base east,inner sep=0pt, outer sep=0pt, scale=  1.00] at ( 37.20, 60.36) {0};

\node[text=drawColor,anchor=base east,inner sep=0pt, outer sep=0pt, scale=  1.00] at ( 37.20, 73.37) {20};

\node[text=drawColor,anchor=base east,inner sep=0pt, outer sep=0pt, scale=  1.00] at ( 37.20, 86.39) {40};

\node[text=drawColor,anchor=base east,inner sep=0pt, outer sep=0pt, scale=  1.00] at ( 37.20, 99.40) {60};

\node[text=drawColor,anchor=base east,inner sep=0pt, outer sep=0pt, scale=  1.00] at ( 37.20,112.41) {80};

\node[text=drawColor,anchor=base east,inner sep=0pt, outer sep=0pt, scale=  1.00] at ( 37.20,125.43) {100};
\end{scope}
\begin{scope}
\path[clip] (  0.00,  0.00) rectangle (260.17,180.67);
\definecolor{drawColor}{RGB}{0,0,0}

\node[text=drawColor,anchor=base,inner sep=0pt, outer sep=0pt, scale=  1.00] at (142.09, 30.60) {Prerequisite test result (dichotomous scoring)};

\node[text=drawColor,rotate= 90.00,anchor=base,inner sep=0pt, outer sep=0pt, scale=  1.00] at ( 15.80, 96.34) {Game score};
\end{scope}
\begin{scope}
\path[clip] (  0.00,  0.00) rectangle (260.17,180.67);
\definecolor{drawColor}{RGB}{0,0,0}

\path[draw=drawColor,line width= 0.4pt,line join=round,line cap=round] ( 49.20, 61.20) --
	(234.97, 61.20) --
	(234.97,131.47) --
	( 49.20,131.47) --
	( 49.20, 61.20);
\end{scope}
\begin{scope}
\path[clip] ( 49.20, 61.20) rectangle (234.97,131.47);
\definecolor{drawColor}{RGB}{0,0,0}

\node[text=drawColor,anchor=base,inner sep=0pt, outer sep=0pt, scale=  1.00] at ( 70.41,121.00) {$\scriptstyle n = 1 $};

\node[text=drawColor,anchor=base,inner sep=0pt, outer sep=0pt, scale=  1.00] at ( 99.08,121.00) {$\scriptstyle n = 0 $};

\node[text=drawColor,anchor=base,inner sep=0pt, outer sep=0pt, scale=  1.00] at (127.75,121.00) {$\scriptstyle n = 3 $};

\node[text=drawColor,anchor=base,inner sep=0pt, outer sep=0pt, scale=  1.00] at (156.42,121.00) {$\scriptstyle n = 9 $};

\node[text=drawColor,anchor=base,inner sep=0pt, outer sep=0pt, scale=  1.00] at (185.09,121.00) {$\scriptstyle n = 18 $};

\node[text=drawColor,anchor=base,inner sep=0pt, outer sep=0pt, scale=  1.00] at (213.76,121.00) {$\scriptstyle n = 16 $};
\end{scope}
\end{tikzpicture}
\end{minipage}%
\begin{minipage}{.5\textwidth}
\begin{tikzpicture}[x=1pt,y=1pt]
\definecolor{fillColor}{RGB}{255,255,255}
\path[use as bounding box,fill=fillColor,fill opacity=0.00] (0,0) rectangle (260.17,180.67);
\begin{scope}
\path[clip] ( 49.20, 61.20) rectangle (234.97,131.47);
\definecolor{drawColor}{RGB}{0,0,0}

\path[draw=drawColor,line width= 1.2pt,line join=round] ( 66.83, 94.06) -- ( 88.33, 94.06);

\path[draw=drawColor,line width= 0.4pt,dash pattern=on 4pt off 4pt ,line join=round,line cap=round] ( 77.58, 63.80) -- ( 77.58, 70.96);

\path[draw=drawColor,line width= 0.4pt,dash pattern=on 4pt off 4pt ,line join=round,line cap=round] ( 77.58,119.11) -- ( 77.58,115.21);

\path[draw=drawColor,line width= 0.4pt,line join=round,line cap=round] ( 72.21, 63.80) -- ( 82.96, 63.80);

\path[draw=drawColor,line width= 0.4pt,line join=round,line cap=round] ( 72.21,119.11) -- ( 82.96,119.11);

\path[draw=drawColor,line width= 0.4pt,line join=round,line cap=round] ( 66.83, 70.96) --
	( 88.33, 70.96) --
	( 88.33,115.21) --
	( 66.83,115.21) --
	( 66.83, 70.96);

\path[draw=drawColor,line width= 1.2pt,line join=round] (109.83, 90.48) -- (131.34, 90.48);

\path[draw=drawColor,line width= 0.4pt,dash pattern=on 4pt off 4pt ,line join=round,line cap=round] (120.58, 69.01) -- (120.58, 80.07);

\path[draw=drawColor,line width= 0.4pt,dash pattern=on 4pt off 4pt ,line join=round,line cap=round] (120.58,128.87) -- (120.58,104.15);

\path[draw=drawColor,line width= 0.4pt,line join=round,line cap=round] (115.21, 69.01) -- (125.96, 69.01);

\path[draw=drawColor,line width= 0.4pt,line join=round,line cap=round] (115.21,128.87) -- (125.96,128.87);

\path[draw=drawColor,line width= 0.4pt,line join=round,line cap=round] (109.83, 80.07) --
	(131.34, 80.07) --
	(131.34,104.15) --
	(109.83,104.15) --
	(109.83, 80.07);

\path[draw=drawColor,line width= 1.2pt,line join=round] (152.84,107.07) -- (174.34,107.07);

\path[draw=drawColor,line width= 0.4pt,dash pattern=on 4pt off 4pt ,line join=round,line cap=round] (163.59, 99.59) -- (163.59,104.80);

\path[draw=drawColor,line width= 0.4pt,dash pattern=on 4pt off 4pt ,line join=round,line cap=round] (163.59,116.51) -- (163.59,113.26);

\path[draw=drawColor,line width= 0.4pt,line join=round,line cap=round] (158.21, 99.59) -- (168.96, 99.59);

\path[draw=drawColor,line width= 0.4pt,line join=round,line cap=round] (158.21,116.51) -- (168.96,116.51);

\path[draw=drawColor,line width= 0.4pt,line join=round,line cap=round] (152.84,104.80) --
	(174.34,104.80) --
	(174.34,113.26) --
	(152.84,113.26) --
	(152.84,104.80);

\path[draw=drawColor,line width= 0.4pt,line join=round,line cap=round] (163.59, 90.48) circle (  2.25);

\end{scope}
\begin{scope}
\path[clip] (  0.00,  0.00) rectangle (260.17,180.67);
\definecolor{drawColor}{RGB}{0,0,0}

\path[draw=drawColor,line width= 0.4pt,line join=round,line cap=round] ( 77.58, 61.20) -- (206.59, 61.20);

\path[draw=drawColor,line width= 0.4pt,line join=round,line cap=round] ( 77.58, 61.20) -- ( 77.58, 55.20);

\path[draw=drawColor,line width= 0.4pt,line join=round,line cap=round] (120.58, 61.20) -- (120.58, 55.20);

\path[draw=drawColor,line width= 0.4pt,line join=round,line cap=round] (163.59, 61.20) -- (163.59, 55.20);

\path[draw=drawColor,line width= 0.4pt,line join=round,line cap=round] (206.59, 61.20) -- (206.59, 55.20);

\node[text=drawColor,anchor=base,inner sep=0pt, outer sep=0pt, scale=  1.00] at ( 77.58, 44.60) {0};

\node[text=drawColor,anchor=base,inner sep=0pt, outer sep=0pt, scale=  1.00] at (120.58, 44.60) {1};

\node[text=drawColor,anchor=base,inner sep=0pt, outer sep=0pt, scale=  1.00] at (163.59, 44.60) {2};

\node[text=drawColor,anchor=base,inner sep=0pt, outer sep=0pt, scale=  1.00] at (206.59, 44.60) {3};

\path[draw=drawColor,line width= 0.4pt,line join=round,line cap=round] ( 49.20, 63.80) -- ( 49.20,128.87);

\path[draw=drawColor,line width= 0.4pt,line join=round,line cap=round] ( 49.20, 63.80) -- ( 43.20, 63.80);

\path[draw=drawColor,line width= 0.4pt,line join=round,line cap=round] ( 49.20, 76.82) -- ( 43.20, 76.82);

\path[draw=drawColor,line width= 0.4pt,line join=round,line cap=round] ( 49.20, 89.83) -- ( 43.20, 89.83);

\path[draw=drawColor,line width= 0.4pt,line join=round,line cap=round] ( 49.20,102.84) -- ( 43.20,102.84);

\path[draw=drawColor,line width= 0.4pt,line join=round,line cap=round] ( 49.20,115.86) -- ( 43.20,115.86);

\path[draw=drawColor,line width= 0.4pt,line join=round,line cap=round] ( 49.20,128.87) -- ( 43.20,128.87);

\node[text=drawColor,anchor=base east,inner sep=0pt, outer sep=0pt, scale=  1.00] at ( 37.20, 60.36) {0};

\node[text=drawColor,anchor=base east,inner sep=0pt, outer sep=0pt, scale=  1.00] at ( 37.20, 73.37) {20};

\node[text=drawColor,anchor=base east,inner sep=0pt, outer sep=0pt, scale=  1.00] at ( 37.20, 86.39) {40};

\node[text=drawColor,anchor=base east,inner sep=0pt, outer sep=0pt, scale=  1.00] at ( 37.20, 99.40) {60};

\node[text=drawColor,anchor=base east,inner sep=0pt, outer sep=0pt, scale=  1.00] at ( 37.20,112.41) {80};

\node[text=drawColor,anchor=base east,inner sep=0pt, outer sep=0pt, scale=  1.00] at ( 37.20,125.43) {100};
\end{scope}
\begin{scope}
\path[clip] (  0.00,  0.00) rectangle (260.17,180.67);
\definecolor{drawColor}{RGB}{0,0,0}

\node[text=drawColor,anchor=base,inner sep=0pt, outer sep=0pt, scale=  1.00] at (142.09, 30.60) {Self-assessment result};

\node[text=drawColor,rotate= 90.00,anchor=base,inner sep=0pt, outer sep=0pt, scale=  1.00] at ( 15.80, 96.34) {Game score};
\end{scope}
\begin{scope}
\path[clip] (  0.00,  0.00) rectangle (260.17,180.67);
\definecolor{drawColor}{RGB}{0,0,0}

\path[draw=drawColor,line width= 0.4pt,line join=round,line cap=round] ( 49.20, 61.20) --
	(234.97, 61.20) --
	(234.97,131.47) --
	( 49.20,131.47) --
	( 49.20, 61.20);
\end{scope}
\begin{scope}
\path[clip] ( 49.20, 61.20) rectangle (234.97,131.47);
\definecolor{drawColor}{RGB}{0,0,0}

\node[text=drawColor,anchor=base,inner sep=0pt, outer sep=0pt, scale=  1.00] at ( 77.58,121.00) {$\scriptstyle n = 14 $};

\node[text=drawColor,anchor=base,inner sep=0pt, outer sep=0pt, scale=  1.00] at (120.58,121.00) {$\scriptstyle n = 23 $};

\node[text=drawColor,anchor=base,inner sep=0pt, outer sep=0pt, scale=  1.00] at (163.59,121.00) {$\scriptstyle n = 10 $};

\node[text=drawColor,anchor=base,inner sep=0pt, outer sep=0pt, scale=  1.00] at (206.59,121.00) {$\scriptstyle n = 0 $};
\end{scope}
\end{tikzpicture}
\end{minipage}%
\vspace*{-1.2cm}
\caption{Boxplots depicting relationships between test score, self-assessment, and game score in G1}
\label{fig:boxplots}
\end{figure*}

\begin{figure*}
\centering
\vspace*{-1.7cm}
\begin{minipage}{.5\textwidth}
\begin{tikzpicture}[x=1pt,y=1pt]
\definecolor{fillColor}{RGB}{255,255,255}
\path[use as bounding box,fill=fillColor,fill opacity=0.00] (0,0) rectangle (260.17,180.67);
\begin{scope}
\path[clip] ( 49.20, 61.20) rectangle (234.97,131.47);
\definecolor{drawColor}{RGB}{0,0,0}
\definecolor{fillColor}{RGB}{0,0,0}

\path[draw=drawColor,line width= 0.4pt,line join=round,line cap=round,fill=fillColor] (193.69, 90.48) circle (  1.5);

\path[draw=drawColor,line width= 0.4pt,line join=round,line cap=round,fill=fillColor] (193.69,100.89) circle (  1.5);

\path[draw=drawColor,line width= 0.4pt,line join=round,line cap=round,fill=fillColor] (193.69,104.80) circle (  1.5);

\path[draw=drawColor,line width= 0.4pt,line join=round,line cap=round,fill=fillColor] (193.69, 99.59) circle (  1.5);

\path[draw=drawColor,line width= 0.4pt,line join=round,line cap=round,fill=fillColor] (228.09,104.80) circle (  1.5);

\path[draw=drawColor,line width= 0.4pt,line join=round,line cap=round,fill=fillColor] (159.29,100.89) circle (  1.5);

\path[draw=drawColor,line width= 0.4pt,line join=round,line cap=round,fill=fillColor] (228.09, 82.02) circle (  1.5);

\path[draw=drawColor,line width= 0.4pt,line join=round,line cap=round,fill=fillColor] (193.69, 84.63) circle (  1.5);

\path[draw=drawColor,line width= 0.4pt,line join=round,line cap=round,fill=fillColor] (193.69,114.56) circle (  1.5);

\path[draw=drawColor,line width= 0.4pt,line join=round,line cap=round,fill=fillColor] (124.88,115.21) circle (  1.5);

\path[draw=drawColor,line width= 0.4pt,line join=round,line cap=round,fill=fillColor] (228.09,111.30) circle (  1.5);

\path[draw=drawColor,line width= 0.4pt,line join=round,line cap=round,fill=fillColor] (193.69, 91.78) circle (  1.5);

\path[draw=drawColor,line width= 0.4pt,line join=round,line cap=round,fill=fillColor] (124.88, 63.80) circle (  1.5);

\path[draw=drawColor,line width= 0.4pt,line join=round,line cap=round,fill=fillColor] (228.09,116.51) circle (  1.5);

\path[draw=drawColor,line width= 0.4pt,line join=round,line cap=round,fill=fillColor] (159.29,104.80) circle (  1.5);

\path[draw=drawColor,line width= 0.4pt,line join=round,line cap=round,fill=fillColor] (228.09,104.80) circle (  1.5);

\path[draw=drawColor,line width= 0.4pt,line join=round,line cap=round,fill=fillColor] (228.09, 69.01) circle (  1.5);

\path[draw=drawColor,line width= 0.4pt,line join=round,line cap=round,fill=fillColor] (228.09,116.51) circle (  1.5);

\path[draw=drawColor,line width= 0.4pt,line join=round,line cap=round,fill=fillColor] (193.69, 89.83) circle (  1.5);

\path[draw=drawColor,line width= 0.4pt,line join=round,line cap=round,fill=fillColor] (228.09,114.56) circle (  1.5);

\path[draw=drawColor,line width= 0.4pt,line join=round,line cap=round,fill=fillColor] (159.29, 98.29) circle (  1.5);

\path[draw=drawColor,line width= 0.4pt,line join=round,line cap=round,fill=fillColor] (159.29,110.65) circle (  1.5);

\path[draw=drawColor,line width= 0.4pt,line join=round,line cap=round,fill=fillColor] (228.09,119.11) circle (  1.5);

\path[draw=drawColor,line width= 0.4pt,line join=round,line cap=round,fill=fillColor] (193.69,116.51) circle (  1.5);

\path[draw=drawColor,line width= 0.4pt,line join=round,line cap=round,fill=fillColor] ( 56.08, 70.96) circle (  1.5);

\path[draw=drawColor,line width= 0.4pt,line join=round,line cap=round,fill=fillColor] (193.69, 80.07) circle (  1.5);

\path[draw=drawColor,line width= 0.4pt,line join=round,line cap=round,fill=fillColor] (159.29, 90.48) circle (  1.5);

\path[draw=drawColor,line width= 0.4pt,line join=round,line cap=round,fill=fillColor] (228.09, 90.48) circle (  1.5);

\path[draw=drawColor,line width= 0.4pt,line join=round,line cap=round,fill=fillColor] (193.69, 90.48) circle (  1.5);

\path[draw=drawColor,line width= 0.4pt,line join=round,line cap=round,fill=fillColor] (193.69, 66.41) circle (  1.5);

\path[draw=drawColor,line width= 0.4pt,line join=round,line cap=round,fill=fillColor] (159.29, 72.26) circle (  1.5);

\path[draw=drawColor,line width= 0.4pt,line join=round,line cap=round,fill=fillColor] (159.29, 67.71) circle (  1.5);

\path[draw=drawColor,line width= 0.4pt,line join=round,line cap=round,fill=fillColor] (193.69, 74.86) circle (  1.5);

\path[draw=drawColor,line width= 0.4pt,line join=round,line cap=round,fill=fillColor] (193.69,103.50) circle (  1.5);

\path[draw=drawColor,line width= 0.4pt,line join=round,line cap=round,fill=fillColor] (193.69, 73.56) circle (  1.5);

\path[draw=drawColor,line width= 0.4pt,line join=round,line cap=round,fill=fillColor] (228.09,114.56) circle (  1.5);

\path[draw=drawColor,line width= 0.4pt,line join=round,line cap=round,fill=fillColor] (193.69,112.60) circle (  1.5);

\path[draw=drawColor,line width= 0.4pt,line join=round,line cap=round,fill=fillColor] (159.29, 87.23) circle (  1.5);

\path[draw=drawColor,line width= 0.4pt,line join=round,line cap=round,fill=fillColor] (228.09,128.87) circle (  1.5);

\path[draw=drawColor,line width= 0.4pt,line join=round,line cap=round,fill=fillColor] (228.09,104.80) circle (  1.5);

\path[draw=drawColor,line width= 0.4pt,line join=round,line cap=round,fill=fillColor] (228.09,115.86) circle (  1.5);

\path[draw=drawColor,line width= 0.4pt,line join=round,line cap=round,fill=fillColor] (228.09,113.26) circle (  1.5);

\path[draw=drawColor,line width= 0.4pt,line join=round,line cap=round,fill=fillColor] (193.69, 89.18) circle (  1.5);

\path[draw=drawColor,line width= 0.4pt,line join=round,line cap=round,fill=fillColor] (124.88, 72.26) circle (  1.5);

\path[draw=drawColor,line width= 0.4pt,line join=round,line cap=round,fill=fillColor] (159.29,109.35) circle (  1.5);

\path[draw=drawColor,line width= 0.4pt,line join=round,line cap=round,fill=fillColor] (228.09, 88.53) circle (  1.5);

\path[draw=drawColor,line width= 0.4pt,line join=round,line cap=round,fill=fillColor] (193.69, 75.52) circle (  1.5);
\end{scope}
\begin{scope}
\path[clip] (  0.00,  0.00) rectangle (260.17,180.67);
\definecolor{drawColor}{RGB}{0,0,0}

\path[draw=drawColor,line width= 0.4pt,line join=round,line cap=round] ( 56.08, 61.20) -- (228.09, 61.20);

\path[draw=drawColor,line width= 0.4pt,line join=round,line cap=round] ( 56.08, 61.20) -- ( 56.08, 55.20);

\path[draw=drawColor,line width= 0.4pt,line join=round,line cap=round] ( 90.48, 61.20) -- ( 90.48, 55.20);

\path[draw=drawColor,line width= 0.4pt,line join=round,line cap=round] (124.88, 61.20) -- (124.88, 55.20);

\path[draw=drawColor,line width= 0.4pt,line join=round,line cap=round] (159.29, 61.20) -- (159.29, 55.20);

\path[draw=drawColor,line width= 0.4pt,line join=round,line cap=round] (193.69, 61.20) -- (193.69, 55.20);

\path[draw=drawColor,line width= 0.4pt,line join=round,line cap=round] (228.09, 61.20) -- (228.09, 55.20);

\node[text=drawColor,anchor=base,inner sep=0pt, outer sep=0pt, scale=  1.00] at ( 56.08, 44.60) {0};

\node[text=drawColor,anchor=base,inner sep=0pt, outer sep=0pt, scale=  1.00] at ( 90.48, 44.60) {1};

\node[text=drawColor,anchor=base,inner sep=0pt, outer sep=0pt, scale=  1.00] at (124.88, 44.60) {2};

\node[text=drawColor,anchor=base,inner sep=0pt, outer sep=0pt, scale=  1.00] at (159.29, 44.60) {3};

\node[text=drawColor,anchor=base,inner sep=0pt, outer sep=0pt, scale=  1.00] at (193.69, 44.60) {4};

\node[text=drawColor,anchor=base,inner sep=0pt, outer sep=0pt, scale=  1.00] at (228.09, 44.60) {5};

\path[draw=drawColor,line width= 0.4pt,line join=round,line cap=round] ( 49.20, 63.80) -- ( 49.20,128.87);

\path[draw=drawColor,line width= 0.4pt,line join=round,line cap=round] ( 49.20, 63.80) -- ( 43.20, 63.80);

\path[draw=drawColor,line width= 0.4pt,line join=round,line cap=round] ( 49.20, 76.82) -- ( 43.20, 76.82);

\path[draw=drawColor,line width= 0.4pt,line join=round,line cap=round] ( 49.20, 89.83) -- ( 43.20, 89.83);

\path[draw=drawColor,line width= 0.4pt,line join=round,line cap=round] ( 49.20,102.84) -- ( 43.20,102.84);

\path[draw=drawColor,line width= 0.4pt,line join=round,line cap=round] ( 49.20,115.86) -- ( 43.20,115.86);

\path[draw=drawColor,line width= 0.4pt,line join=round,line cap=round] ( 49.20,128.87) -- ( 43.20,128.87);

\node[text=drawColor,anchor=base east,inner sep=0pt, outer sep=0pt, scale=  1.00] at ( 37.20, 60.36) {0};

\node[text=drawColor,anchor=base east,inner sep=0pt, outer sep=0pt, scale=  1.00] at ( 37.20, 73.37) {20};

\node[text=drawColor,anchor=base east,inner sep=0pt, outer sep=0pt, scale=  1.00] at ( 37.20, 86.39) {40};

\node[text=drawColor,anchor=base east,inner sep=0pt, outer sep=0pt, scale=  1.00] at ( 37.20, 99.40) {60};

\node[text=drawColor,anchor=base east,inner sep=0pt, outer sep=0pt, scale=  1.00] at ( 37.20,112.41) {80};

\node[text=drawColor,anchor=base east,inner sep=0pt, outer sep=0pt, scale=  1.00] at ( 37.20,125.43) {100};

\path[draw=drawColor,line width= 0.4pt,line join=round,line cap=round] ( 49.20, 61.20) --
	(234.97, 61.20) --
	(234.97,131.47) --
	( 49.20,131.47) --
	( 49.20, 61.20);
\end{scope}
\begin{scope}
\path[clip] (  0.00,  0.00) rectangle (260.17,180.67);
\definecolor{drawColor}{RGB}{0,0,0}

\node[text=drawColor,anchor=base,inner sep=0pt, outer sep=0pt, scale=  1.00] at (142.09, 30.60) {Prerequisite test result (dichotomous scoring)};

\node[text=drawColor,rotate= 90.00,anchor=base,inner sep=0pt, outer sep=0pt, scale=  1.00] at ( 15.80, 96.34) {Game score};
\end{scope}
\begin{scope}
\path[clip] ( 49.20, 61.20) rectangle (234.97,131.47);
\definecolor{drawColor}{RGB}{0,0,255}

\path[draw=drawColor,line width= 0.8pt,line join=round,line cap=round] ( 49.20, 68.30) -- (234.97,104.63);
\end{scope}
\end{tikzpicture}
\end{minipage}%
\begin{minipage}{.5\textwidth}
\begin{tikzpicture}[x=1pt,y=1pt]
\definecolor{fillColor}{RGB}{255,255,255}
\path[use as bounding box,fill=fillColor,fill opacity=0.00] (0,0) rectangle (260.17,180.67);
\begin{scope}
\path[clip] ( 49.20, 61.20) rectangle (234.97,131.47);
\definecolor{drawColor}{RGB}{0,0,0}
\definecolor{fillColor}{RGB}{0,0,0}

\path[draw=drawColor,line width= 0.4pt,line join=round,line cap=round,fill=fillColor] (181.89, 90.48) circle (  1.5);

\path[draw=drawColor,line width= 0.4pt,line join=round,line cap=round,fill=fillColor] (197.13,100.89) circle (  1.5);

\path[draw=drawColor,line width= 0.4pt,line join=round,line cap=round,fill=fillColor] (192.71,104.80) circle (  1.5);

\path[draw=drawColor,line width= 0.4pt,line join=round,line cap=round,fill=fillColor] (185.33, 99.59) circle (  1.5);

\path[draw=drawColor,line width= 0.4pt,line join=round,line cap=round,fill=fillColor] (216.79,104.80) circle (  1.5);

\path[draw=drawColor,line width= 0.4pt,line join=round,line cap=round,fill=fillColor] (138.65,100.89) circle (  1.5);

\path[draw=drawColor,line width= 0.4pt,line join=round,line cap=round,fill=fillColor] (154.86, 82.02) circle (  1.5);

\path[draw=drawColor,line width= 0.4pt,line join=round,line cap=round,fill=fillColor] (170.59, 84.63) circle (  1.5);

\path[draw=drawColor,line width= 0.4pt,line join=round,line cap=round,fill=fillColor] (193.20,114.56) circle (  1.5);

\path[draw=drawColor,line width= 0.4pt,line join=round,line cap=round,fill=fillColor] (154.86,115.21) circle (  1.5);

\path[draw=drawColor,line width= 0.4pt,line join=round,line cap=round,fill=fillColor] (211.87,111.30) circle (  1.5);

\path[draw=drawColor,line width= 0.4pt,line join=round,line cap=round,fill=fillColor] (201.55, 91.78) circle (  1.5);

\path[draw=drawColor,line width= 0.4pt,line join=round,line cap=round,fill=fillColor] (133.24, 63.80) circle (  1.5);

\path[draw=drawColor,line width= 0.4pt,line join=round,line cap=round,fill=fillColor] (177.96, 89.83) circle (  1.5);

\path[draw=drawColor,line width= 0.4pt,line join=round,line cap=round,fill=fillColor] ( 60.50, 70.96) circle (  1.5);

\path[draw=drawColor,line width= 0.4pt,line join=round,line cap=round,fill=fillColor] (173.54, 80.07) circle (  1.5);

\path[draw=drawColor,line width= 0.4pt,line join=round,line cap=round,fill=fillColor] (177.47, 90.48) circle (  1.5);

\path[draw=drawColor,line width= 0.4pt,line join=round,line cap=round,fill=fillColor] (210.40, 90.48) circle (  1.5);

\path[draw=drawColor,line width= 0.4pt,line join=round,line cap=round,fill=fillColor] (181.40, 90.48) circle (  1.5);

\path[draw=drawColor,line width= 0.4pt,line join=round,line cap=round,fill=fillColor] (186.32, 66.41) circle (  1.5);

\path[draw=drawColor,line width= 0.4pt,line join=round,line cap=round,fill=fillColor] (151.92, 72.26) circle (  1.5);

\path[draw=drawColor,line width= 0.4pt,line join=round,line cap=round,fill=fillColor] (167.64, 67.71) circle (  1.5);

\path[draw=drawColor,line width= 0.4pt,line join=round,line cap=round,fill=fillColor] (181.89, 74.86) circle (  1.5);

\path[draw=drawColor,line width= 0.4pt,line join=round,line cap=round,fill=fillColor] (187.79,103.50) circle (  1.5);

\path[draw=drawColor,line width= 0.4pt,line join=round,line cap=round,fill=fillColor] (174.03, 73.56) circle (  1.5);

\path[draw=drawColor,line width= 0.4pt,line join=round,line cap=round,fill=fillColor] (199.59,114.56) circle (  1.5);

\path[draw=drawColor,line width= 0.4pt,line join=round,line cap=round,fill=fillColor] (176.98,112.60) circle (  1.5);

\path[draw=drawColor,line width= 0.4pt,line join=round,line cap=round,fill=fillColor] (164.20, 87.23) circle (  1.5);

\path[draw=drawColor,line width= 0.4pt,line join=round,line cap=round,fill=fillColor] (223.18,128.87) circle (  1.5);

\path[draw=drawColor,line width= 0.4pt,line join=round,line cap=round,fill=fillColor] (216.79,104.80) circle (  1.5);

\path[draw=drawColor,line width= 0.4pt,line join=round,line cap=round,fill=fillColor] (218.26,115.86) circle (  1.5);

\path[draw=drawColor,line width= 0.4pt,line join=round,line cap=round,fill=fillColor] (223.18,113.26) circle (  1.5);

\path[draw=drawColor,line width= 0.4pt,line join=round,line cap=round,fill=fillColor] (173.05, 89.18) circle (  1.5);

\path[draw=drawColor,line width= 0.4pt,line join=round,line cap=round,fill=fillColor] (139.14, 72.26) circle (  1.5);

\path[draw=drawColor,line width= 0.4pt,line join=round,line cap=round,fill=fillColor] (155.36,109.35) circle (  1.5);

\path[draw=drawColor,line width= 0.4pt,line join=round,line cap=round,fill=fillColor] (205.48, 88.53) circle (  1.5);

\path[draw=drawColor,line width= 0.4pt,line join=round,line cap=round,fill=fillColor] (174.03, 75.52) circle (  1.5);
\end{scope}
\begin{scope}
\path[clip] (  0.00,  0.00) rectangle (260.17,180.67);
\definecolor{drawColor}{RGB}{0,0,0}

\path[draw=drawColor,line width= 0.4pt,line join=round,line cap=round] ( 56.08, 61.20) -- (228.09, 61.20);

\path[draw=drawColor,line width= 0.4pt,line join=round,line cap=round] ( 56.08, 61.20) -- ( 56.08, 55.20);

\path[draw=drawColor,line width= 0.4pt,line join=round,line cap=round] ( 80.65, 61.20) -- ( 80.65, 55.20);

\path[draw=drawColor,line width= 0.4pt,line join=round,line cap=round] (105.23, 61.20) -- (105.23, 55.20);

\path[draw=drawColor,line width= 0.4pt,line join=round,line cap=round] (129.80, 61.20) -- (129.80, 55.20);

\path[draw=drawColor,line width= 0.4pt,line join=round,line cap=round] (154.37, 61.20) -- (154.37, 55.20);

\path[draw=drawColor,line width= 0.4pt,line join=round,line cap=round] (178.95, 61.20) -- (178.95, 55.20);

\path[draw=drawColor,line width= 0.4pt,line join=round,line cap=round] (203.52, 61.20) -- (203.52, 55.20);

\path[draw=drawColor,line width= 0.4pt,line join=round,line cap=round] (228.09, 61.20) -- (228.09, 55.20);

\node[text=drawColor,anchor=base,inner sep=0pt, outer sep=0pt, scale=  1.00] at ( 56.08, 44.60) {-100};

\node[text=drawColor,anchor=base,inner sep=0pt, outer sep=0pt, scale=  1.00] at ( 80.65, 44.60) {-50};

\node[text=drawColor,anchor=base,inner sep=0pt, outer sep=0pt, scale=  1.00] at (105.23, 44.60) {0};

\node[text=drawColor,anchor=base,inner sep=0pt, outer sep=0pt, scale=  1.00] at (129.80, 44.60) {50};

\node[text=drawColor,anchor=base,inner sep=0pt, outer sep=0pt, scale=  1.00] at (154.37, 44.60) {100};

\node[text=drawColor,anchor=base,inner sep=0pt, outer sep=0pt, scale=  1.00] at (178.95, 44.60) {150};

\node[text=drawColor,anchor=base,inner sep=0pt, outer sep=0pt, scale=  1.00] at (203.52, 44.60) {200};

\node[text=drawColor,anchor=base,inner sep=0pt, outer sep=0pt, scale=  1.00] at (228.09, 44.60) {250};

\path[draw=drawColor,line width= 0.4pt,line join=round,line cap=round] ( 49.20, 63.80) -- ( 49.20,128.87);

\path[draw=drawColor,line width= 0.4pt,line join=round,line cap=round] ( 49.20, 63.80) -- ( 43.20, 63.80);

\path[draw=drawColor,line width= 0.4pt,line join=round,line cap=round] ( 49.20, 76.82) -- ( 43.20, 76.82);

\path[draw=drawColor,line width= 0.4pt,line join=round,line cap=round] ( 49.20, 89.83) -- ( 43.20, 89.83);

\path[draw=drawColor,line width= 0.4pt,line join=round,line cap=round] ( 49.20,102.84) -- ( 43.20,102.84);

\path[draw=drawColor,line width= 0.4pt,line join=round,line cap=round] ( 49.20,115.86) -- ( 43.20,115.86);

\path[draw=drawColor,line width= 0.4pt,line join=round,line cap=round] ( 49.20,128.87) -- ( 43.20,128.87);

\node[text=drawColor,anchor=base east,inner sep=0pt, outer sep=0pt, scale=  1.00] at ( 37.20, 60.36) {0};

\node[text=drawColor,anchor=base east,inner sep=0pt, outer sep=0pt, scale=  1.00] at ( 37.20, 73.37) {20};

\node[text=drawColor,anchor=base east,inner sep=0pt, outer sep=0pt, scale=  1.00] at ( 37.20, 86.39) {40};

\node[text=drawColor,anchor=base east,inner sep=0pt, outer sep=0pt, scale=  1.00] at ( 37.20, 99.40) {60};

\node[text=drawColor,anchor=base east,inner sep=0pt, outer sep=0pt, scale=  1.00] at ( 37.20,112.41) {80};

\node[text=drawColor,anchor=base east,inner sep=0pt, outer sep=0pt, scale=  1.00] at ( 37.20,125.43) {100};

\path[draw=drawColor,line width= 0.4pt,line join=round,line cap=round] ( 49.20, 61.20) --
	(234.97, 61.20) --
	(234.97,131.47) --
	( 49.20,131.47) --
	( 49.20, 61.20);
\end{scope}
\begin{scope}
\path[clip] (  0.00,  0.00) rectangle (260.17,180.67);
\definecolor{drawColor}{RGB}{0,0,0}

\node[text=drawColor,anchor=base,inner sep=0pt, outer sep=0pt, scale=  1.00] at (142.09, 30.60) {Prerequisite test result (confidence scoring)};

\node[text=drawColor,rotate= 90.00,anchor=base,inner sep=0pt, outer sep=0pt, scale=  1.00] at ( 15.80, 96.34) {Game score};
\end{scope}
\begin{scope}
\path[clip] ( 49.20, 61.20) rectangle (234.97,131.47);
\definecolor{drawColor}{RGB}{0,0,255}

\path[draw=drawColor,line width= 0.8pt,line join=round,line cap=round] ( 49.20, 54.00) -- (234.97,109.95);
\end{scope}
\end{tikzpicture}
\end{minipage}%
\vspace*{-1.5cm}
\caption{Linear regression models describing game score by pretest for G1}
\label{fig:models}
\end{figure*}

Table~\ref{table:stats} reports the examined variables and descriptive statistics of the collected data for G1, which are further detailed in Figure~\ref{fig:boxplots}. The boxplots show distributions of the game score ($T$) grouped by the dichotomous quiz score ($P_d$) and the input from the self-assessment ($S$). The boxplots showing the distributions of finished levels ($L$) are almost identical considering patterns of the medians, and thus were omitted. Also, statistically significant ($p \leq 0.02$) Pearson and Spearman correlations (ranging from $0.36$ to $0.60$) were reported between the score and pretest and between levels completed and pretest (both scoring methods in both cases). Finally, there was strong evidence that all performance predictors and skill descriptors negatively correlate with how difficult the game is perceived ($p \leq 0.02$, coefficients ranging from $-0.37$ to $-0.52$).

Next, we modeled the learners' skill (dependent variables $T$ and $L$) using independent variables $P_d$, $P_c$, and $S$. Three different sets of regression analyses were performed. First, we used the data of the 47 participants playing G1, ignoring the two extra pretest questions given in session 4. Second, we used only the data of the 12 participants playing G1, using the result of the 7-question pretest. Third, we used the data of the 20 participants playing G2.

Table \ref{table:models} reports the regression models of the first set of analyses. Statistically significant fits were computed for the score or level prediction based on the dichotomously scored pretest ($R^2 = 0.17$ or $R^2 = 0.14$, $p \leq 0.01$). An even more promising relationship emerged when incorporating confidence testing ($R^2 = 0.31$ or $R^2 = 0.24$, $p \leq 0.004$). The second set of analyses on the subset of players and two extra questions yielded almost identical results to those on the full sample of players from G1. The coefficients in the remaining models did not show statistical significance, and neither did they in other models for G1 nor in any model for G2.

The two best fits for score prediction are graphed in Figure~\ref{fig:models}. These plots show a certain degree of linearity. The other two plots for predicting completed levels were largely similar and were omitted to conserve space. The regression diagnostic plots (residual plots and Q-Q plots) confirm that the assumptions of homoscedasticity and multivariate normality were met. Based on the leverage plot, the player scoring 0 points in the pretest was identified as an outlier (Cook's distance $> 0.5$). However, the removal of the data point did not significantly influence the models (the $R^2$ changed by $\pm 0.01$, and the p-value remained up to $0.01$). Thus, we decided to keep the data point in the sample.

\begin{figure*}
\centering
\includegraphics[width=\linewidth]{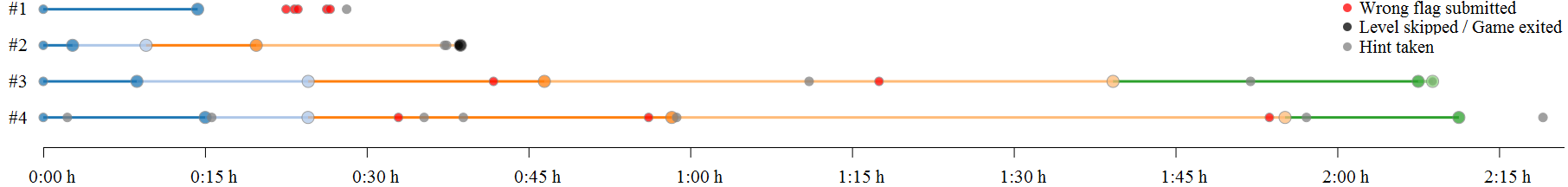}
\caption{Game events of selected individual players in G1, distributed over time. Each line represents the actions of one player. Finished levels are marked as colored line segments: dark blue displays level 1, light blue level 2, and so on.}
\label{fig:events}
\end{figure*}

\section{Discussion} \label{sec:discussion}

\subsection{Quantitative view}

Returning to the research questions posed in Section 1, the models accurately predicted the learners' total game score and completed levels when employing prerequisite testing, regardless of the scoring method used, for the game G1. It is interesting that confidence testing revealed guesswork: 8 players randomly guessed at least one correct answer. It also showed misconceptions, since 7 players were quite or absolutely sure about at least one incorrect answer. Most importantly, the confidence testing improved the models, as the value of $R^2$ almost doubled compared to the dichotomous pretest.

Considering G1, the players from the fourth session exhibited similar results as the whole when using the two extra pretest questions. However, the success was not reproduced in G2. We attribute this to the fact that two of the questions were easy for the players to answer. Moreover, due to unclear formulation of the third question we decided to accept even the answers we originally deemed incorrect. As a result, 16 out of 20 players had at least three questions correct, preventing the models from differentiating between them.

To complement the discussion of the regression models, we examined medians in boxplots depicting prerequisite test results (Figure~\ref{fig:boxplots}). On the one hand, the medians do not show a linear trend, as they are not in ascending order. On the other hand, the players who achieved the highest quiz scores also had the highest medians of the game score and finished levels. Overall, the results confirmed the intuition that players with a high score from the pretest would often perform better in the game compared to the others. The reverse was also true. 

However, the regression models employing self-assessment were not statistically significant, and their $R^2$ values were below $0.07$. The models seem unlikely to reach statistical significance even if the sample size increased. The conclusion that the self-assessment is an unreliable skill predictor is consistent with the previous findings by Mirkovic~et.~al~\cite{mirkovic}, who used a similar self-assessment scale. One possible explanation is that experience with using in-game tools is not a key factor in deciding learners' readiness. Most of the players had worked with a Linux Terminal before the game, thus were able to discover and understand the application of other command-line tools for themselves. Another plausible justification for the poor self-assessment results is that the scale has only four values and three items, which yields data that is too coarse.

\subsection{Qualitative view}

It can be argued that applying statistical tests and using regression models on a relatively small sample might bias the quantitative results. Therefore, the actions of individual players were further explored from a time perspective (see Figure~\ref{fig:events}). Several notable anomalies were identified in G1 sessions and are addressed below\footnote{All the players are referred to as males, even if their gender is unknown.}.

Player \#1, who we nicknamed ``the dropout'', reached the full score (5 points) in the dichotomous pretest but got frustrated as early as in level 2. Over the course of less than 6 minutes, he attempted 5 wrong flags, took a hint, and stopped playing. As a result, his game score was only 8 points.

Player \#2 (``the achiever'') followed the same pattern as ``the dropout''; at the beginning, he seemed like a competent learner but got frustrated with the game. ``The achiever'' scored 4 points in the dichotomous pretest and solved the first three levels quickly. Shortly before the indicative time limit for the fourth level ran out, he took both available hints and then prematurely exited the game. As a result, he scored a, below-average, 43 game points. However, he later reported not knowing that the time limit was only informative and had no impact on the game. Instead, the player thought that if the time runs out, he cannot play anymore, which annoyed him and caused him to quit the game.

We hypothesize that ``the achiever'' and ``the dropout'' had possessed the necessary prerequisites for finishing the game. However, they were thwarted by ambiguous game mechanics or design, by insufficient attention paid to the rules, or by some other reason. Due to these unanticipated situations, the dataset includes players scoring well in the pretest but poorly in terms of game score or levels completed. This might have introduced noise in the regression models, as similar misunderstandings could have influenced other players' results.

Another interesting case is player \#3 (``the determined one''), who scored 0 in the self-assessment and 2 points in the pretest. Still, he completed 5 levels and scored 79 points in the game: one of the best results in the sample. The player used only 2 hints and attempted only 2 incorrect flags in total, all in the later phases of the game. The time spent in the levels was rather long. This is reflected in the post-game feedback, where he rated the game as hard (4). Although the player did not possess theoretical knowledge from the pretest, his determination allowed him to perform very well.

Finally, the player \#4 (``the practitioner'') scored 0 in the self-assessment and 3 points in the pretest. However, by taking some hints, using trial and error, and given enough time he was able to complete 5 levels and score 72 game points, which is a good result. This player, like ``the determined one'', might not have had the theoretical background, but was still able to solve the practical tasks.

These case anomalies show that some unanticipated aspects influence players' performance. An arising challenge is recognizing and deeply understanding all factors that contribute to a successful game. We believe that solving this challenge is essential for designing a useful diagnostic assessment and the whole game.

\vspace*{-0.3cm}
\subsection{Addressing the limitations}

Despite using a well-established framework for question-writing~\cite{beatty} and following best practices of assessment design, we were confronted by three main challenges of prerequisite testing. The first is calibrating the test to predict the possession of skills most relevant to the game. While the players often performed well in the quiz, no one finished the last level of G1. It seems that theoretical knowledge might not be enough for succeeding in practical tasks. 

The second challenge was the limited time frame for assessing a participant. It is impractical and discouraging to perform a lengthy examination when the learners are eager to play the game. Both the test and self-assessment combined were designed to take 8 minutes at most,  yet were perceived by some of the players as an inconvenience. The third challenge is embedding the pretest in the game. For further experiments, we propose designing and implementing pretests, and, by extension, cybersecurity games differently. Inspired by the results of Lee at al.~\cite{lee}, who report positive effects of assessments in educational games, we propose two main improvements. One is dissolving the assessments into the story of the game. Compared to using questionnaires, which distract the players and shift them into a ``testing mode'', in-game tests are more engaging~\cite{lee}. They also allow the use of more assessment questions, which, in turn, brings more validity and reliability to the results. This approach necessitates another improvement in the design of the game itself. Individual levels can be created such that each has only one particular learning outcome. Appropriate prerequisites can be tested before or during that level.

While the experiment proved a link between prerequisite testing and players' performance in the game G1, the generalizability of the results might be questionable, as the model for G2 was not statistically significant. One possible explanation is the dependence on the particular game and its scoring method. Ultimately, successful diagnostic assessment largely depends on the quality of game design. This work attempted to prove the validity of the proposed prerequisite test based, to some extent, on its relationship to the game. However, if the game scoring mechanism or individual levels are poorly designed, this can invalidate the pretest. Therefore, we underline the need for careful consideration of educational game design. Another explanation is that the amount and content of questions in pretest 2 were not sufficient to differentiate the players' skill. Nevertheless, adding more questions in pretest 1 did not seem to bring more validity to the results, as they reflected the results of the analysis performed with the subset of questions.

Finally, the self-assessment had only three items with four distinct values, which was too coarse an input for the linear regression. Introducing questions asking for the frequency of use or applicability to a particular task could increase the reliability of results.

\vspace*{-0.3cm}
\section{Conclusions} \label{sec:conclusions}

We performed an experimental study investigating the predictive value of prerequisite tests and self-assessment for identifying learners' readiness before playing a cybersecurity educational game. The analysis of game events and information provided by players showed that only a knowledge quiz, and not self-assessment, can model game score and completed levels. The models based on the pretests significantly improved if the quiz contained confidence assessment. However, educators have to pay special attention to the selection and formulation of quiz questions since they fundamentally affect the accuracy of the model.

The major contribution of this pioneering attempt is the new insights provided into hands-on cybersecurity education, which has not been widely researched. This work also motivated the development of an open-source tool for visualizing generic game events over time~\cite{uhlar} (see Figure~\ref{fig:events}). This tool allowed discovering important patterns that would otherwise stay hidden.

In our future work, we will focus on investigating the means of diagnostic assessment into the game story and structure. Since the players rated the games as educational, practical, and interesting, we believe that active learning in cybersecurity is worthy of both security practitioners' and educators' attention. We also encourage fellow researchers to experiment with diagnostic assessment in other domains than cybersecurity.

\vspace{-0.2cm}
\begin{acks}
This research was supported by the Security Research Programme of the Czech Republic 2015-2020 (BV III/1--VS) granted by the \grantsponsor{cz-interior}{Ministry of the Interior of the Czech Republic}{http://www.mvcr.cz/bezpecnostni-vyzkum.aspx} under No.~\grantnum{cz-interior}{VI20162019014} -- Simulation, detection, and mitigation of cyber threats endangering critical infrastructure. The analysis of the experiment results would not have been possible without a visualization tool developed by Juraj Uhl\'{a}r. Finally, we thank our colleagues and anonymous reviewers for their helpful comments and suggestions.
\end{acks}

\bibliographystyle{ACM-Reference-Format}
\bibliography{references}

\end{document}